\documentstyle[12pt]{article}
\begin{document}
\centerline{\bf TEST OF THE INVERSE SQUARE LAW THROUGH}
\centerline{\bf PRECESSION OF ORBITS} \vskip5mm \centerline{N.I.
Kolosnitsyn and V.N.Melnikov} \centerline{\it Center for
Gravitation and Fundamental Metrology,VNIIMS} \centerline{\it and
Institute of Gravitation and Cosmology,} \centerline{\it Peoples'
Friendship University of Russia,} \centerline{\it 3-1, M.Ulyanovoi
Str., 119313, Moscow, Russia}\centerline{\rm e-mail:
melnikov@rgs.phys.msu.su}
 \vskip5mm
\centerline{\bf Abstract} \vskip5mm Using precession of orbits due
to non-Newtonian interaction between two celestial bodies and
modern tracking data of satellites, planets and a pulsar we obtain
new more precise
 limits on possible Yukawa-type deviations from the
 Newton law in planets (satellites)
  radii ranges.
 \vskip5mm
{\bf 1. Non-Newtonian interactions, or range variations of
G.}\vskip5mm
Many modified theories of gravity and unified
theories \cite{Mel2,Mel,3} predict
 some deviations from the Newton law (inverse square law, ISL) or
 composition-dependent
 violations of the Equivalence Principle (EP) due to appearance of
 additional fields or
 new possible massive particles (partners) \cite{4,5}.
 Experimental data limit the
 existence of these effects at nearly all ranges at some level,
 but the most poor
 data are available at less than millimeter and also at meters
 and hundreds of meters
 ranges. The only positive result of existence of such deviation in
 the range of 20 to
 500 m was obtained by Achilli et al. using an energy storage plant
 experiment with
 gravimeters. They found the deviation from the Newton law with
 the Yukawa potential
 strength $\alpha$ between 0.13 and 0.25. This result
 contradicts other experimental
 data and needs to be verified in similar or other
 independent experiments, probably
 in space ones \cite{5,solv,10}.

In the Einstein theory G is a true constant. But, if we think that
G may vary with time \cite{3,4}, then, from a relativistic point
of view, it may vary with distance as well.  In GR massless
gravitons are mediators of the gravitational interaction, they
obey second-order differential equations and interact with matter
with a constant strength G. If any of these requirements is
violated or we consider quantum corrections to the classical
theory, we come in general to deviations from the Newton law with
range (or to generalization of GR).

In \cite{5} several classes of such theories were analysed:

1. Theories with massive gravitons like bimetric ones or theories with
a $\Lambda$-term.

2. Theories with an effective gravitational constant like the
general scalar-tensor
ones \cite{4}.

3. Theories with torsion.

4. Theories with higher derivatives (4th-order equations etc.), where massive modes
in a propagator appear leading to short-range additional forces.

5. More elaborated theories with other mediators besides gravitons (partners),
like supergravity, superstrings, M-theory etc.

6. Theories with nonlinearities induced by any known physical
interactions (Born-Infeld etc.), where an effective mass appears.

7. Phenomenological models, where the detailed mechanism of deviation is not known
(fifth or other force).

In all these theories some effective or real masses appear leading
to Yukawa-type deviation from the Newton law, characterized by
strength $\alpha$ and range $\lambda$ or to a power law deviation
(see \cite{5}).

There exist some model-dependant estimations of these forces. The
most well-known one belongs to Scherk (1979) from supergravity
where the graviton is accompanied by a spin-1 partner
(graviphoton) leading to an additional repulsion.  Other models
were suggested by Moody and Wilczek (1984) -- introduction of a
pseudo-scalar particle -- leading to an additional attraction
between macro-bodies with the range $2\cdot10^{-4}$ cm $< \lambda
< 20$ cm and strength $\alpha$ from $1$ to $10^{-10}$ in this
range. Another supersymmetric model was elaborated by Fayet (1986,
1990), where a spin-1 partner of a massive graviton gives an
additional repulsion in the range of the order $10^{3}$ km and
$\alpha$ of the order $10^{-13}$.

A scalar field to adjust $\Lambda$ was introduced also by S.
Weinberg in 1989, with a mass smaller than $10^{-3}$ ${\rm
eV}/c^{2}$, or a range greater than 0.1 mm.  One more variant was
suggested by Peccei, Sola and Wetterich (1987) leading to
additional attraction with a range smaller than 10 km. Some
p-brane models also predict non-Newtonian additional interactions,
in particular in the mm range, what is intensively discussed
nowadays in the hierarchy and braneworld models. About PPN
parameters for multidimensional models with p-branes see
\cite{TR}.

In this paper we consider limits on a possible Yukawa type
additional interaction in the ranges  of the order of planets
(satellites) distances from the Sun (the Earth) using the
precession method and satellites plus planets tracking data.
\vskip 5mm
 {\bf 2. Basic formulas.}
\vskip5mm
 As it is  known, in two bodies problem an orbit of a
celestial body is closed for only two interaction potentials
\cite{1}. They are:

(1) the Newtonian potential:  $U \sim 1/r$ and

(2) $U \sim r^2$.

In other cases the orbit is not closed and a pericenter precession
is observed. In particular, the deviation from the Newton law due
to the Yukawa additional interaction  \vskip3mm
\begin{equation}
U^{\prime}=\frac{Gm_1 m_2}{r}\alpha\exp(-r/\lambda) \label{1}
\end{equation}             %%(1)
\vskip3mm
 entails a precession of an
orbit.

 In a general case the precession magnitude due to a small
perturbation, described
 by a potential $\delta U$, is equal to (see \cite{1})
 \vskip3mm
 \begin{equation}\label{2}
 \delta\varphi=\frac{\partial}{\partial
 M}\left(\frac{2m}{M}\int\limits_0^\pi r^2 \delta U d\varphi\right)
 \end{equation}             %%2
\vskip3mm

Integration is done over a "non-perturbed" trajectory. Here $m_1$
is a mass of one celestial body, $m_2$ is the mass of another
celestial body, $M=mr^2 \dot \varphi$  is the integral of motion
(the angular momentum), $m=m_1 m_2/(m_1+m_2)$ is the reduced mass,

$\delta U=\alpha (Gm_1 m_2/r)\exp(-r/\lambda)$.

The "non-perturbed" trajectory is described by the expressions:
\vskip3mm
\begin{equation}
r=\frac{p}{1+e \cos\varphi},\,\;\;  e^2=1+\frac{2EM^2}{m(Gm_1
m_2)^2},\,\;\; p=\frac{M^2}{mGm_1 m_2}=a(1-e^2).\label{3}
\end{equation}  %%(3)
\vskip3mm where $e$ is an eccentricity, $a$ is a semi-major axis.

 After differentiating the right-hand side of Eq.(2) with
respect to M we obtain \vskip3mm
\begin{equation}
\delta\varphi=2mGm_1
m_2\alpha\left\{-\frac{1}{M^2}\int\limits_0^\pi
re^{-r/\lambda}d\varphi+
\frac{1}{M}\int\limits_0^\pi\left(1-\frac{r}{\lambda}\right)e^{-r/\lambda}
\frac{\partial r}{\partial M}d\varphi\right\}
 \label{4}   %%4
\end{equation}
\vskip3mm

Taking into account  Eq. (3) we get
\vskip3mm
\begin{equation}
\frac{\partial r}{\partial
M}=\frac{r}{M}\left[1+\frac{e+\cos\varphi}{e(1+e\cos\varphi)}\right]\label{5}
\end{equation}%%%5)
\vskip3mm
 After substitution of  (5) in (4) we obtain
 \vskip3mm
\begin{equation}
\delta\varphi=2\alpha\frac{mGm_1 m_2}{M^2}\int\limits_0^\pi
re^{-r/\lambda}\left[-\frac{e+\cos\varphi}{e(1+
e\cos\varphi)}+\frac{r}{\lambda}\left(1+\frac{e+\cos\varphi}{e(1+
e\cos\varphi)}\right)\right]d\varphi \label{6} %%(6)
\end{equation}
\vskip3mm
 As a result, we have
 \vskip3mm
\begin{equation}
\delta\varphi=\alpha\frac{2}{e}\int\limits_0^\pi
\frac{\exp(-r/\lambda)}{(1+e\cos\varphi)^2}\left\{
\frac{r}{\lambda}\left[2e+(1+e^2)\cos\varphi\right]-(e+\cos\varphi)\right\}d\varphi
\label{7} %%(7)
\end{equation}
\vskip3mm
 where
 \vskip3mm
\begin{equation}
\frac{r}{\lambda}=\frac{a}{\lambda}\frac{1-e^2}{(1+e\cos\varphi)}.
\label{8} %%(8)
\end{equation}
\vskip5mm
 {\bf 3. Estimation of effects.}
\vskip5mm
     Using Eq. (7) and proposed data on $\delta\varphi$ error for the
     LAGEOS Satellites,
     the inner planets, Moon and the binary pulser B1913+16, we calculated
     curves $\alpha(\lambda)$,
     which determine  a bound in the plane ($\alpha,\lambda$)  between two domains,
     where the Yukawa interaction (a new nonnewtonian force) is forbidden by experiment
     and where it is not. The sensitivities to Yukawa interactions - the united curve
     $\alpha(\lambda)$ described by Eq.(7) - are shown in Fig.1 as the domain 1998
     for the
     parameter $\lambda$ in the range from $1\cdot 10^6$ m to $1\cdot 10^{13}$ m.

     We took  data for LAGEOS and LAGEOS II from \cite{3a} related to the relativistic
     Lense-Thirring precession. The data contain information on classical (Newtonian)
     and relativistic precession. The last one equals to $\sim 57$ milliarcseconds per year
     for LAGEOS II. The error for test of
     the Lense-Thirring effect is equal to $20\%$. Therefore, we took the error
     $\delta\varphi = 11.4$ milliarcseconds per year. The LAGEOS II eccentricity $e = 0.014$.
     Using the data we obtained $\alpha_{min} = 1.38\cdot10^{-11}$
     for $\lambda= 6.081\cdot10^6$ m. Our result is in a very good agreement with the
     estimation for $\alpha \sim 1\cdot 10^{-11}$ obtained by Iorio (see
     \cite{A}.

     For the Moon there are high precision data of Laser Ranging (see, for example \cite{4a}).
     In Ref. \cite{4a}  geodetic precession (first calculated by de Sitter), equaled to 19 ms
     per year, was determined with
     the error $0.9\%$. The absolute error was equal to $1,71\cdot 10^{-4} "/y$. We used
     this  estimation and obtained $\alpha_{min}=
     3.64\cdot 10^{-11}$ for $ \lambda= 1,92\cdot 10^8$ m.

     Very precise information on the Mercury precession is contained in the article of
     Pitjeva \cite{6a}. Using radar observations of Mercury for the time span 1964-1989
     and taking its topography Pitjeva has estimated systematic errors for the perihelion
     motion and a new random error for this motion as
     $0.052"/cy$. From this it follows that $\alpha_{min} = 3.57\cdot 10^{-10}$ for
     $(\lambda = 2.89\cdot 10^{10}$ m).

      The most precise data on the Mars precession are obtained from observations
      of the landing spacecraft Pathfinder (see \cite{7} ). Pitjeva has analysed joint
      observation data of Viking and Pathfinder missions and has
      obtained more precise estimation for precession of the Mars
      \cite{12}.
      However a sensitivity
      to the Yukawa force in this case is very small and  the sensitivity curve for Mercury dominates.

      We also used data for the binary pulsar PSR 1913+16. The pulsar has an orbit
      with the high eccentricity $e = 0.617$ and a big precession $4,23^o /y\pm 0,000007^o
      /y$.
      Using parameters of the pulsar in \cite{8} we calculated $\alpha_{min}=6.409\cdot
      10^{-11}$ for $\lambda= 7.515\cdot 10^8$ m.

      Fig.1 shows our data (LAGEOS, MOON, PSR1913, MERCURY) as the new experimental
      constraint on the coupling
      parameter $\alpha$ as a function of the parameter $\lambda$ taken from
      the reference \cite{9}.
      \vskip5mm
{ \bf 4. Conclusions} \vskip5mm
 Here, we presented some new limits
on possible deviations from the inverse square law using planets
and satellite tracking data. They improve the existing ones by
several orders at satellite ($\sim 6\cdot10^6$ m) and planets
 distance from the Sun ($\sim 1\cdot10^{13}$ m)
ranges. Some proposed space projects (such as SEE) make possible
to improve these estimations to five orders at a meter range (see
\cite{10} and \cite{2} ). Limits at less than mm ranges see in
\cite{11}. 
\end{document}